\documentclass[12pt,preprint,natbib,iop]{emulateapj}    
\usepackage{graphicx,epsfig,hyperref}   

\lefthead{van der Wel et al.}
\righthead{Compact $z\sim 2$ disks}
\slugcomment{The Astrophysical Journal, submitted}
\begin{document}

\newcommand{\kms}{\>{\rm km}\,{\rm s}^{-1}}
\newcommand{\reff}{R_{\rm{eff}}}
\newcommand{\msol}{M_{\odot}}

\title{The Majority of Compact Massive Galaxies at $z\sim 2$ are Disk
  Dominated}

\author{Arjen van der Wel\altaffilmark{1}, Hans-Walter
  Rix\altaffilmark{1}, Stijn Wuyts\altaffilmark{2}, Elizabeth
  J.~McGrath\altaffilmark{3}, Anton M.~Koekemoer\altaffilmark{4}, \\ Eric
  F.~Bell\altaffilmark{5}, Bradford P.~Holden\altaffilmark{3}, Aday
  R.~Robaina\altaffilmark{6}, Daniel H.~McIntosh\altaffilmark{7}}

\altaffiltext{1}{Max-Planck Institut f\"ur Astronomie, K\"onigstuhl
  17, D-69117, Heidelberg, Germany; vdwel@mpia.de}

\altaffiltext{2}{Max-Planck-Institut f\"ur Extraterrestrische Physik,
  Giessenbachstrasse, D-85748 Garching, Germany}

\altaffiltext{3}{University of California Observatories/Lick
  Observatory, University of California, Santa Cruz, CA 95064, USA}

\altaffiltext{4}{Space Telescope Science Institute, 3700 San Martin
  Drive, Baltimore, MD 21218, USA}

\altaffiltext{5}{Department of Astronomy, University of Michigan, 500
  Church Street, Ann Arbor, Michigan, 48109, USA}

\altaffiltext{6}{Institut de Ci\`encies del Cosmos, University of
  Barcelona, Mart\'e i Franqu\`es, 1 E-08028, Barcelona, Spain}

\altaffiltext{7}{Department of Physics, University of Missouri-Kansas
  City, Kansas City, MO 64110, USA}

\begin{abstract}
  We investigate the stellar structure of massive, quiescent galaxies
  at $z\sim 2$, based on \textit{Hubble Space Telescope}/WFC3 imaging
  from the Early Release Science program.  Our sample of 14 galaxies
  has stellar masses of $M_* > 10^{10.8}~\msol$ and photometric
  redshifts of $1.5 < z < 2.5$.  In agreement with previous work,
  their half-light radii are $<2$ kpc, much smaller than equally
  massive galaxies in the present-day universe.  A significant subset
  of the sample appears highly flattened in projection, which implies,
  considering viewing angle statistics, that a significant fraction of
  the galaxies in our sample have pronounced disks.  This is
  corroborated by two-dimensional surface brightness profile fits.  We
  estimate that $65\%\pm 15\%$ of the population of massive, quiescent
  $z\sim 2$ galaxies are disk-dominated.  The median disk scale length
  is 1.5 kpc, substantially smaller than the disks of equally massive
  galaxies in the present-day universe.  Our results provide strong
  observational evidence that the much-discussed ultra-dense
  high-redshift galaxies should generally be thought of as disk-like
  stellar systems with the majority of stars formed from gas that had
  time to settle into a disk.
\end{abstract}

\section{Introduction}
Structural evolution in the population of non-starforming (quiescent)
galaxies has been observed and extensively discussed over the past few
years. Such evolution has become an important ingredient in our
description of evolutionary processes, and is used to constrain the
theoretical framework for the formation of massive galaxies.  The
small sizes ($R_{\rm{eff}} < 2$~kpc) of massive ($M\sim
10^{11}~\msol$) quiescent galaxies at $z\sim 2$ have received
particular attention, both in terms of explaining their formation and
their subsequent evolution \citep[e.g.,][]{trujillo04, daddi05,
  papovich05, khochfar06b, trujillo06a, zirm07, toft07, vandokkum08b,
  vanderwel08c, mcgrath08, buitrago08, damjanov09, vanderwel09a,
  saracco09, hopkins09a, vandokkum10a, cassata10, mancini10,
  szomoru10}.

The structural properties and surface brightness profiles of these
compact, massive $z\sim 2$ galaxies, beyond their small sizes, can
help us understand the formation scenario for these remarkable objects
and their subsequent evolution: to produce such compact stellar
systems, highly dissipative formation mechanisms have been proposed
\citep[e.g.,][]{hopkins09a, wuyts10}.  Although it is not clear
whether stars have time to settle into a disk in such a scenario, this
may imply that these objects should be rotating.  On the other hand,
given their high stellar masses, clustering properties
\citep[e.g.,][]{quadri07, hartley10}, and number densities
\citep[e.g.,][]{vandokkum10a}, these objects must evolve into very
massive, large, pressure-supported, bulge-dominated galaxies in the
present-day universe.

In this paper we explore the internal structures and surface
brightness profiles of massive, quiescent galaxies at $z\sim 2$,
utilizing \textit{Hubble Space Telescope (HST)}/WFC3 imaging, taken as
part of the Early Release Science (ERS) program.  We show that these
galaxies are predominantly disk-like and describe the differences with
their likely present-day descendants.  From this, a consistent
narrative emerges in which highly dissipative events produce compact,
disk-dominated stellar systems, which subsequently experience a series
of merger events, simultaneously explaining the growth in size and
mass, and the lack of prominent disks in their descendants.

\section{Sample Selection and Data Description}
FIREWORKS \citep{wuyts08} provides a multi-wavelength catalog, from
ground-based \textit{U} band to \textit{Spitzer} $24\mu$m, and
photometric redshifts.  Stellar masses are derived for all sources
with a minimum signal-to-noise ratio in the \textit{K} band of 5, and
photometric data points from at least four other bands.  Following the
procedure outlined by \citet{marchesini09}, stellar-mass-to-light
ratios are estimated from modeling the spectral energy distribution
with the \citet{bruzual03} stellar population synthesis model for
solar metallicity and exponentially declining star formation rates.
We adopt the \citet{kroupa01} stellar initial mass function, and the
assumed cosmology is $(\Omega_{\rm{M}},~\Omega_{\Lambda},~h) =
(0.3,~0.7,~0.7)$.

We select all galaxies with photometric redshifts $1.5 < z < 2.5$ and
stellar masses $M>10^{10.8}~\msol$, at which the catalog is complete.
Based on their rest-frame $r-z$ and $u-r$ colors we select those
galaxies which are quiescent (see Figure 1): galaxies with blue $u-r$
colors are excluded on the basis of the directly observed young
stellar populations; galaxies with red $r-z$ colors are excluded based
on the red color long-ward of the $4000\AA$/Balmer break, indicative
of reddening by dust.  Such a technique was first used by
\citet{wuyts07} and, subsequently, for a much larger sample by
\citet{williams09a}.  Whereas these authors use the $U-V$ and $V-J$
color combination, our $u-r$ and $r-z$ color combination is
essentially equally effective (B.~P.~Holden et al.~, 2011, in
preparation).

Sixteen of these galaxies are in the Chandra Deep Field-South
WFC3/ERS2 region \citep{windhorst10}, for which an independent
reduction of the 10 pointings (5000s total) into an F160W mosaic was
carried out by A.~M.~Koekemoer et al.~(2011, in preparation) using
Multidrizzle \citep{koekemoer02}.

\begin{figure}[t]
\epsscale{1.2} 
\plotone{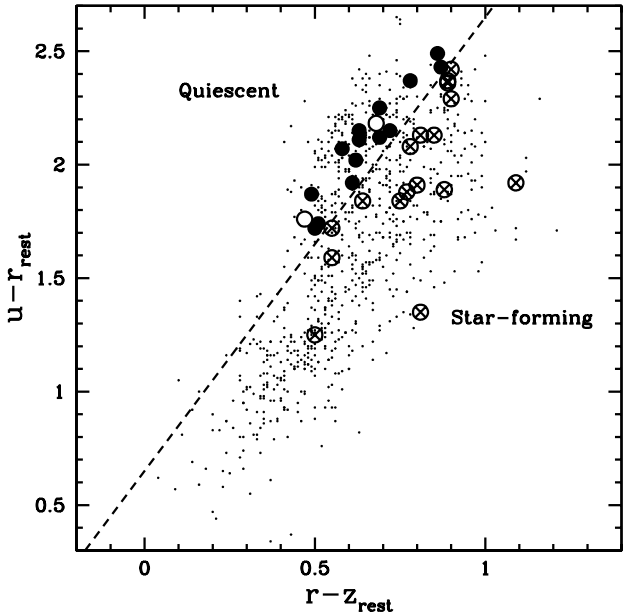}
\caption{ Rest-frame $u-r$ vs.~$r-z$ color for galaxies drawn from the
  FIREWORKS catalog \citep{wuyts08}.  Dots represent galaxies in the
  redshift range $1<z<2$ and with stellar masses
  $M>10^{10}~M_{\odot}$, shown to demonstrate that galaxies separate
  into a star-forming sequence (below the dashed line), along which
  the extinction increases toward the top right, and a quiescent
  sequence \citep[see][for a more clearly defined bi-modal
  distribution]{williams09a}.  Large symbols represent galaxies with
  stellar masses $M>10^{10.8}~M_{\odot}$ in our redshift range of
  interest ($1.5<z<2.5$) and within the \textit{HST}/WFC3 ERS mosaic.
  We select the 14 objects (indicated by filled circles) upward from
  the dashed line as our sample of massive, quiescent $z\sim 2$
  galaxies.  The objects indicated with open circles with crosses are
  considered star forming and not included in our sample.  Two
  galaxies, indicated by open circles, that satisfy the color-criteria
  are excluded from the analysis (see the text).}
\label{f1}
\end{figure}

Two of the 16 galaxies are excluded from further analysis: one is a
merger with strong tidal features; the other is located at a small
projected distance from a very bright foreground galaxy, which
precludes accurate photometry and analysis.

We subject the images of the 14 remaining galaxies to the
Lucy-Robertson deconvolution algorithm (``lucy'' in IRAF), iterating
16 times, using a point spread function (PSF) created with TinyTim
\citep{krist95}, taking into account the dither pattern of the
observations and the Multidrizzle data processing.  The reconstruction
brings out small-scale ($\sim$800 pc), high surface brightness
features.

We show the original and deconvolved images in Figure 2.  Structural
parameters are inferred from fitting single-component Sersic models
with GALFIT \citep{peng02} to the original images, using the same PSF
as described above.  All 14 galaxies, which have a median stellar mass
of $10^{11.1}~M_{\odot}$, have circularized half-light radii smaller
than 2 kpc (the median is 1.2 kpc), in agreement with previous
studies, a median Sersic index of 2.5, and a median axis ratio of
0.67, almost identical to that of the sample of \citet{vandokkum08b}.

We also determine, in the same manner, the structural parameters of
the star-forming galaxies, that is, those below the dashed line in
Figure 1.  Their median Sersic index and axis ratio are $n=1.5$ and
$b/a=0.53$, respectively.  Their sizes vary with distance to the
dashed line in Figure 1: those close to the line have sizes $1.5-2$
kpc, comparable to the galaxies in our quiescent sample, while those
far from the line are larger ($4-5$ kpc).  This implies that our
results do not depend on the exact criterion for quiescence.

\begin{figure*}[t]
\epsscale{1.1} 
\plotone{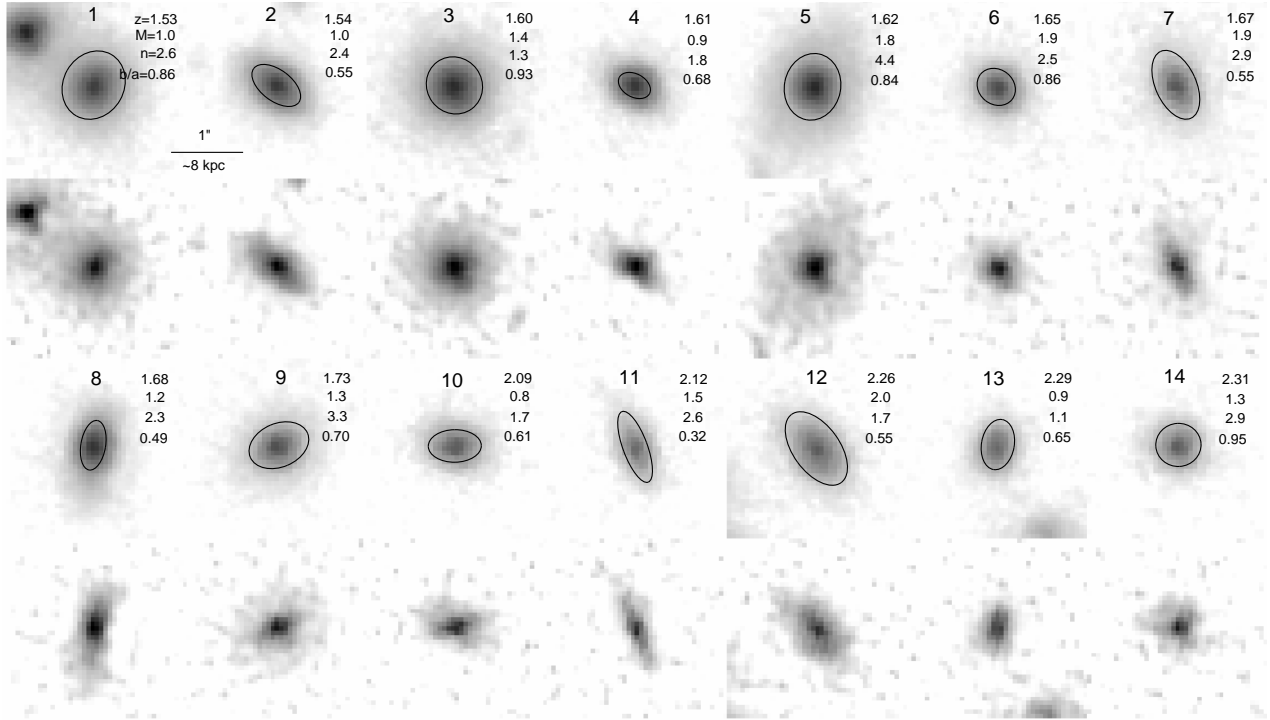}
\caption{ \textit{HST/WFC3} F160W imaging of our sample of massive,
  quiescent $z\sim 2$ galaxies (all with $H_{\rm{F160W}}<23$).  The
  deconvolved images are shown directly below the original images.
  Photometric redshifts, stellar masses (in units of $10^{11}~\msol$),
  Sersic indices, and axis ratios from one-component profile fits, as
  well as the scale of the images, are indicated.  Ellipses indicate
  best-fitting axis ratios and sizes from GALFIT -- the area of the
  ellipse corresponds to that of a circle with a radius that is twice
  the circularized half-light radius.  All galaxies are clearly
  resolved and many are flattened in projection, indicative of a
  disk-like stellar structure.}
\label{f2}
\end{figure*}

\section{The Disk-Like Nature of Massive $z=2$ Galaxies}

A significant fraction of the galaxies shown in Figure 2 are strongly
flattened in projection -- the galaxies 2, 7, 8, 11, and 12 have
best-fitting axis ratios $b/a \lesssim 0.5$, indicative of a disk-like
structure.  Viewed more face-on, this would be difficult to
recognize. Assuming approximately random viewing angles, this implies
that a significant fraction -- at face value a majority -- of massive,
quiescent $z\sim 2$ are disk-dominated.

To follow up on the evidence based on the axis ratio distribution, we
examine the disk-like nature of the individual galaxies by performing
two-component fits with GALFIT on the original images, using the same
PSF as before. The positions of the two components are kept fixed at
the position of the one-component fit, and the Sersic index is fixed at
$n=1$ for the disk-like component.  All other parameters are left
free.  In Figure 3 we show 4 examples of the two-component fits and
point out the similarity between the two-component model fits and the
deconvolved images shown in Figure 2.

We explore the uncertainty in the two-component fits by artificially
doubling the rms in the background by adding Gaussian noise, and by
using two different PSF models: a PSF that is constructed from bright
stars in the same mosaic, and a simpler TinyTim-based PSF.  These
alternative PSFs are known to be a less accurate representation of the
true PSF than our preferred, default PSF.

For four out of 14 galaxies (nos.~2, 10, 11, 12) we reproduce the
disk-dominated two-component fits regardless of the adopted PSF model
and the artificially increased noise level.  For another three
galaxies (nos.~4, 5, 8) with disk-dominated two-component fits from
our preferred PSF and original noise level, the fitting parameters
change with PSF choice and noise level to the extent that we consider
the evidence for the disk-dominated nature of these objects as
tentative.

For these 4+3 galaxies either the $\chi^2$ value of the two-component
fit is significantly lower than that of the one-component fit, or the
one-component fit is disk-like itself (with Sersic index
$n\lesssim2$).  For the other 7 galaxies the fitting could not
directly ascertain whether they are disk-dominated.

To estimate the disk-dominated fraction among massive, quiescent
$z\sim 2$ galaxies, we combine the results from the quantitative
two-component fits with the visual impression reflected in Figure 2
(mostly flatness), which are independent of each other.  All four
galaxies with robust disk-dominated two-component fits are flat in
projection (with $b/a \leq 0.6$).  We assign those objects weights of
0.9 (90\% probability that these galaxies are, in fact,
disk-dominated).  We assign weight 0.7 to those 3 galaxies with more
tentative quantitative evidence.  These weights/probabilities should
not be considered precise and quantitative, but serve as a proxy for
our level of confidence that these galaxies are disk-dominated.

If we conservatively assume that none of the other seven galaxies are
disk dominated (i.e., they have weight 0), then we infer that
$40\%\pm15\%$ of the population of massive, quiescent $z\sim 2$
galaxies is disk dominated.  This number and its uncertainty include
the weights as specified above and the uncertainty due to the small
sample size.

However, some of the 7 non-classified galaxies, for example, 7 and
13, have small axis ratios.  Therefore, it is not reasonable to
interpret the lack of quantitative evidence for dominant disks as
evidence for the absence of such disks.  If we assign weight 0.5 to
those objects, we infer that $65\%\pm15\%$ of the population of
massive, quiescent $z\sim 2$ galaxies is disk-dominated.

We have not attempted a more formal inversion of the intrinsic axis
ratios as our sample is too small to de-project the observed axis
ratio distribution without making overly restrictive assumptions about
the underlying intrinsic axis ratio distribution.  Qualitatively,
however, the general flatness of the galaxies in our sample is
consistent with a population with disks of finite intrinsic thickness
($\sim 0.25$).  We stress that simply considering the observation that
5 out of 14 galaxies have axis ratios $b/a \leq 0.55$ leads to a
similar conclusion that $\sim 50\%-100\%$ of the galaxies in the
sample must be intrinsically thin, that is, disk-like: face-on
counterparts of the observed edge-on disk-like galaxies must exist as
well.

We note that while very prolate systems can also lead to frequent
small projected axis ratios, we consider this possibility unlikely:
the absence of very prolate self-gravitating systems at lower
redshifts suggests that nature does not produce such objects, and this
should be independent of cosmic time.

Based on examining the also available F125W and F098N imaging of our
sample, we are confident that our inferences from the light profiles
directly translate into information about the stellar mass
distribution.  The F125W and F160W images (roughly corresponding to
\textit{B} and \textit{V} in the rest frame) are very similar in
appearance, and all galaxies are much fainter in the bluer F098N
passband (rest-frame \textit{U}).  We show color images for four
disk-dominated galaxies in Figure 3; these illustrate that strong
color gradients are absent, which implies that the disks are not
strongly star forming.\footnote{None of these examples have 24$\mu$m
  counterparts that signify star formation rates in excess of $\sim
  20\msol \rm{yr^{-1}}$, which is less than the average past star
  formation rate for these galaxies, qualifying them as quiescent.} In
other words, the light distribution must be similar to the stellar
mass distribution.

The major axis exponential scale length of the disks of the galaxies
we consider as disk-dominated ranges from 1 to 4 kpc, with a median of
1.5 kpc; this is substantially smaller than the scale lengths of disks
in similarly massive galaxies in the present-day universe, which is
$\sim 4$~kpc ($\log(R_{\rm{d}}/\rm{kpc}) = 0.6 \pm 0.1$, where 0.1 is
the $1\sigma$ scatter, \citet{fathi10a, fathi10b}).  The median scale
length for our sample deviates by $\sim3\sigma$, which implies that it
is unlikely that most of these disk-like structures will survive up
until the present day.

Several studies have before measured the size evolution of disk-like
and bulge-like galaxies based on their Sersic indices
\citep{trujillo06a, buitrago08}, but only for stellar mass selected
samples that include star-forming galaxies. \citet{stockton06},
\citet{mcgrath08}, and \citet{stockton08} reported the existence of
passive, massive galaxies at $z>1.5$ with disk-like surface brightness
profiles, showing that a mix of morphological properties exists among
passive, high-redshift galaxies.  However, the population of compact
galaxies reported by, for example, \citet{zirm07} and
\citet{vandokkum08b} have not been identified with disk-dominated
morphologies, but are generally thought of as early-type, bulge-like
structures.  Although \citet{vandokkum08b} mention the disk-like
nature of some of the galaxies in their sample, they did not consider
the implications for the overall population statistics.

Our findings suggest that the majority of compact, massive, quiescent
$z\sim 2$ galaxies are disk-dominated, typically hosting stellar disks
with scale radii that are several times smaller than present-day
disks.

\begin{figure*}[t]
\begin{center}
\epsscale{1.8} 
\plottwo{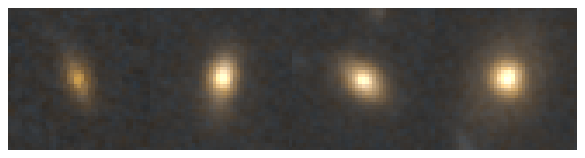}{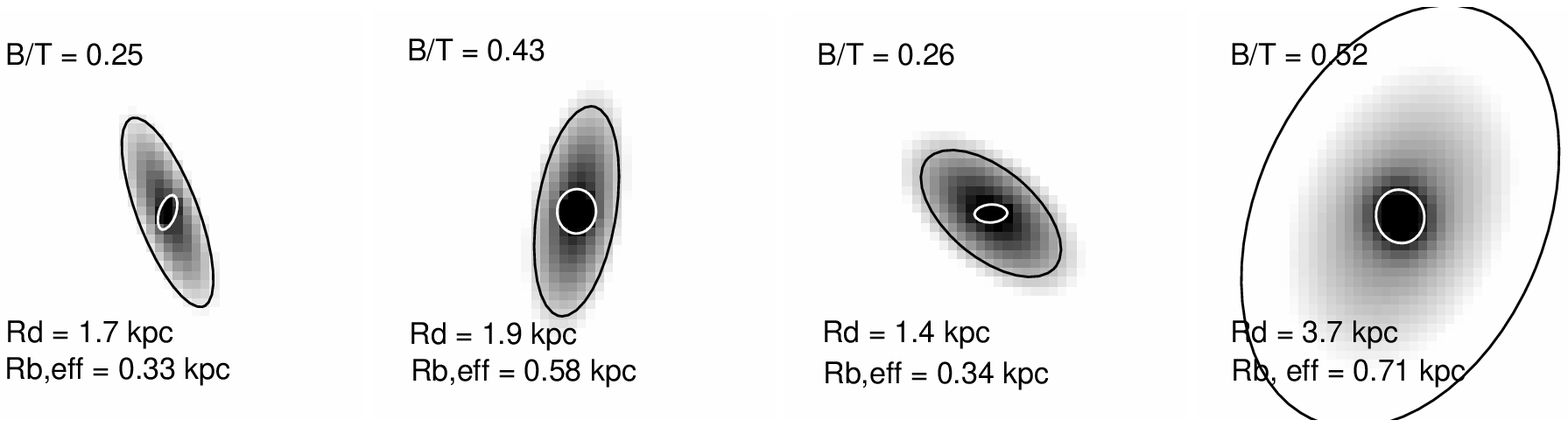}
\caption{ Top: F098N+F160W color composites for galaxies 11, 8, 2, and
  5 from Figure 2, ordered by axis ratio.  These four examples are
  chosen because of their flatness, with the exception of no.~5, which
  appears to have a compact bulge-like component surrounded by a more
  extended, disk-like component.  There is no strong indications for
  color gradients, suggesting that the disk-components of these
  galaxies are not strongly star forming.  Bottom: two-component model
  fits (without PSF smearing) for the same galaxies.  The white and
  black ellipses indicate twice the size of the half-light ellipses
  for bulge-like and disk-like components, respectively.  $B/T$ is the
  ratio of the light in the model for the bulge-like component and the
  light of the models for the two components combined.  ``Rd'' is the
  exponential scale length as measured along the major axis of the
  disk-like component, which we calculate by dividing the semimajor
  axis of the ``half-light ellipse'' by 1.6.  ``Rb,eff'' is the
  circularized half-light radius of the bulge-like component.}
\label{f3}
\end{center}
\end{figure*}

\section{Formation and Evolution}
The scale lengths of the stellar disks in quiescent $z\sim 2$ galaxies
are, on average, $\sim 3$ times smaller than the scale lengths of
similarly massive present-day stellar disks (see Section 3).  This is
consistent with the zeroth-order theoretical expectation: the size and
angular momentum of a $z\sim 2$ dark matter halo are on average $\sim
3$ times smaller than those of a similarly massive present-day dark
matter halo.  The assumption here is that the correspondence between a
disk and its halo in terms of relative mass, size, and angular
momentum does not evolve \citep{mo99}.  While these assumptions are
obviously not strictly valid, it is still not unexpected that stellar
disks are several times smaller at $z\sim 2$ than at the present.  In
addition, the global compactness of quiescent galaxies at $z\sim 2$ is
further boosted by the presence of compact, central components --
these galaxies are not pure disks.

Simulations of gas-rich mergers produce descendants that are broadly
reminiscent of the observed galaxies, most particularly in the sense
that these simulated merger remnants have substantial rotation, and
must therefore be disk-like \citep{wuyts10}.  Moreover, the
simulations produce dense cores, and more extended outer parts.
However, there are important quantitative differences.  The merger
remnants are entirely unlike exponential disks -- they have very high
Sersic indices, $n>4$.  This is the result of overly dominant central
compact components.  Gradual gas accretion to produce a disk, combined
with the occasional (minor) merger event to produce a modest central
component, may explain the described structural properties of
quiescent $z\sim 2$ galaxies.  Generally speaking, their formation
mechanism must be sufficiently gentle and slow as to allow gas to
settle onto a disk before converting into stars.

However, we note that the quiescent galaxies are not simply $z\sim 2$
star-forming galaxies of which the star formation rate has been
reduced from $100-1000~\msol~\rm{yr}^{-1}$ \citep{forster09} to
$\lesssim 20~\msol~\rm{yr}^{-1}$.  Although massive star-forming
galaxies at $z\sim 2$ are also disk-like in their structural
parameters \citep[see Section 3 and, e.g.,][]{labbe03, forster10} as
well as in their dynamical properties \citep[e.g.,][]{forster09,
  genzel10}, their disks are substantially larger (with scale lengths
of $\sim 3$ kpc) than the quiescent disks described here and only
somewhat smaller than present-day disks.  The explanation for this may
be the age difference between the quiescent and star-forming galaxies:
the star-forming galaxies are undergoing a major, possibly first,
episode of significant growth, while the quiescent galaxies must have
undergone such a phase at an earlier epoch, implying smaller sizes
\citep[see also][]{franx08, vanderwel09a}.  The correlation between
color and size for star-forming galaxies (Section 2) is consistent
with this picture.

Let us now consider the evolutionary path between the epoch of
observation, $z\sim 2$, and the present day.  In a hierarchical
$\Lambda$CDM universe, the descendants of $\sim10^{11}~\msol$~$z\sim
2$ galaxies are super-$L^*$ galaxies with masses
$>(2-3)\times10^{11}~\msol$.  This argument is bolstered by the
observed clustering of massive $z\sim 2$ galaxies in comparison with
the clustering properties of very massive galaxies in the present-day
universe \citep{quadri07, hartley10}.  A simple, model-independent
argument to the same effect is that the comoving number density of
$z\sim 2$ galaxies with stellar masses $\sim 10^{11}~\msol$ is the
same as that of present-day galaxies with stellar masses $\sim 3\times
10^{11}~\msol$ \citep{vandokkum10a}.  That is, these disk-like $z\sim
2$ galaxies are \textit{not} progenitors of Milky Way type galaxies at
the present day.

Thus, we have compact, disk-dominated galaxies at $z\sim 2$, the
present-day descendants of which are $2-3$ times more massive, are
$\sim5$ times larger in half-light radius, and almost never have
prominent stellar disks \citep{vanderwel09b}.  Clearly, the structure
of these galaxies has evolved dramatically over the past 10 Gyr.
Growth in mass through merging by the relatively modest factor of two
or three compares well with observed merger rates
\citep[e.g.,][]{robaina10} and model expectations \citep{hopkins10}.
The expected merger trees consist of a mix of frequent minor
mergers/accretion events and rare major mergers with low overall gas
fractions ($\lesssim 2$~per galaxy since $z=2$).  The former are
generally thought to provide the most efficient mechanism to explain
size evolution \citep[e.g.,][]{bezanson09, vanderwel09a}.  Indeed, the
gradual buildup of the outer parts associated with such accretion
events has been observed \citep{vandokkum10a}.  To destroy a massive
stellar disk, major merging is more efficient, although a sequence of
many minor accretion events can have the same effect \citep{naab99,
  bournaud07}.  Overall, a consistent narrative is emerging in which
merging and accretion explain the growth in size and change in
structure of massive, passive galaxies over time.

Further support for the direct link between the disk-like galaxies at
$z\sim 2$ and the pressure-supported, massive elliptical galaxies in
the present-day universe is provided by the comparable stellar
densities of the $z\sim 2$ galaxies and the cores of present-day
ellipticals \citep{bezanson09, hopkins09a, vandokkum10a}.  Dynamical
modeling of very massive nearby ellipticals has revealed that whereas
the global rotation rate is small, the majority of the stars, even in
the inner parts, are on disk-like orbits \citep[see][, for an
example]{vandenbosch08}.  This may be the archaeological remnant of
the disk-like nature of its progenitors.

\acknowledgements{The authors thank the referee, Matthew Bershady, for
  helpful suggestions that helped improve the paper, and Kambiz Fathi
  for sharing his disk scale length measurements and comments on the
  manuscript.}

\bibliographystyle{apj}

\end{document}